\input{aipcheck}

\documentclass[
    ,final            
  ]
  {aipproc}

\layoutstyle{8x11single}
\def\Journal#1#2#3#4{{#1} {\bf #2}, #3 (#4)}

\def\RNC{\em Rivista Nuovo Cimento}

\def\NIMA{{\em Nucl. Instrum. Methods} A}

\def\PLB{{\em Phys. Lett.}  B}
\def\PRL{\em Phys. Rev. Lett.}
\def\PRD{{\em Phys. Rev.} D}

\def\GaC{\em Gravitation and Cosmology}
\def\GaCS{{\em Gravitation and Cosmology} Supplement}

\def\JETPL{\em JETP Lett.}
\def\PAN{\em Phys.Atom.Nucl.}
\def\CQG{\em Class. Quantum Grav.}
\def\APJ{\em Astrophys. J.}
\def\SCI{\em Science}
\def\MPLA{{\em Mod. Phys. Lett.}  A}
\def\IJTP{\em Int. J. Theor. Phys.}
\def\NJP{\em New J. of Phys.}
\def\JHEP{\em JHEP}
\def\BWP{\em Bled Workshops in Physics}
\def\EPHJ{\em Eur.Phys.J}
\def\s{{\,\rm s}}
\def\g{{\,\rm g}}
\def\eV{\,{\rm eV}}
\def\keV{\,{\rm keV}}
\def\MeV{\,{\rm MeV}}
\def\GeV{\,{\rm GeV}}
\def\TeV{\,{\rm TeV}}
\def\sv{\left<\sigma v\right>}
\def\({\left(}
\def\){\right)}
\def\cm{{\,\rm cm}}

\def\kpc{{\,\rm kpc}}
\def\beq{\begin{equation}}
\def\eeq{\end{equation}}
\def\bea{\begin{eqnarray}}
\def\eea{\end{eqnarray}}

\begin{document}

\title{The puzzles of dark matter searches}

\classification{12.60.Cn,98.90.+s,12.60.Nz,14.60.Hi,26.35.+c,36.90.+f,03.65.Ge}
\keywords      {elementary particles, nuclear reactions, nucleosynthesis, abundances,
dark matter, early universe, large-scale structure of universe}

\author{Maxim Yu. Khlopov}{
  address={APC laboratory 10, rue Alice Domon et L\'eonie Duquet \\75205
Paris Cedex 13, France}
  ,altaddress={Centre for Cosmoparticle Physics "Cosmion" 115409 Moscow, Russia;\\ Moscow Engineering Physics Institute (National Nuclear Research University), 115409 Moscow, Russia}
}

\begin{abstract}
 Positive results of dark matter searches in DAMA/NaI and DAMA/LIBRA experiments, being put together with negative results of other groups, can imply nontrivial particle physics solutions for cosmological dark matter. Stable particles with charge -2 bind with primordial helium in O-helium "atoms" (OHe), representing a specific
Warmer than Cold nuclear-interacting form of dark matter. Slowed down in the
terrestrial matter, OHe is elusive for direct methods of underground
Dark matter detection like those used in CDMS experiment, but its
reactions with nuclei can lead to annual variations of energy
release in the interval of energy 2-6 keV in DAMA/NaI and DAMA/LIBRA
experiments. Schrodinger equation for system of nucleus and OHe is considered and reduced to an equation of relative motion in a spherically symmetrical potential well, formed by the Yukawa tail of nuclear scalar isoscalar attraction potential, acting on He beyond the nucleus, and dipole Coulomb repulsion between the nucleus and OHe at distances from the nuclear surface, smaller than the size of OHe. The values of coupling strength and mass of meson, mediating scalar isoscalar nuclear potential, are rather uncertain. Within these uncertainties we find a narrow window of these parameters, at which  the sodium and/or iodine nuclei have a few keV binding energy with OHe. The concentration of OHe in the matter of underground detectors is adjusted to the incoming flux of cosmic O-helium at the timescale less than few minutes. Therefore the rate of radiative capture of Na and/or I by OHe should experience annual modulations. Transitions to more energetic levels of Na+OHe (I+OHe) system imply tunneling through dipole Coulomb barrier that leads to suppression of annual modulation of events with MeV-tens MeV energy release in the correspondence with the results of DAMA experiments. The proposed explanation inevitably leads to prediction of abundance of anomalous Na (and/or I) corresponding to the signal, observed by DAMA. At nuclear parameters, reproducing DAMA results, the energy release predicted for detectors with chemical content other than NaI differ in the most cases from the one in DAMA detector. In particular, it is shown that in the case of CDMS the energy of OHe-germanium bound state is beyond the range of 2-6 keV and its formation should not lead to ionization in the energy interval of DAMA signal.
\end{abstract}

\maketitle


\section{Introduction}

 The widely shared belief is that the dark matter, corresponding to
$25\%$ of the total cosmological density, is nonbaryonic and
consists of new stable particles. One can formulate the set of
conditions under which new particles can be considered as candidates
to dark matter (see e.g. \cite{book,Cosmoarcheology,Bled07} for
review and reference): they should be stable, saturate the measured
dark matter density and decouple from plasma and radiation at least
before the beginning of matter dominated stage. The easiest way to
satisfy these conditions is to involve neutral weakly interacting
particles. However it is not the only particle physics solution for
the dark matter problem. In the composite dark matter scenarios new
stable particles can have electric charge, but escape experimental
discovery, because they are hidden in atom-like states maintaining
dark matter of the modern Universe.

It offers new solutions for the
physical nature of the cosmological dark matter. The main problem
for these solutions is to suppress the abundance of positively
charged species bound with ordinary electrons, which behave as
anomalous isotopes of hydrogen or helium. This problem is
unresolvable, if the model predicts stable particles with charge -1,
as it is the case for tera-electrons \cite{Glashow,Fargion:2005xz}.
To avoid anomalous isotopes overproduction, stable particles with
charge -1 should be absent, so that stable negatively charged
particles should have charge -2 only.

Elementary particle frames for heavy stable -2 charged species are provided by:
(a) stable "antibaryons" $\bar U \bar U \bar U$ formed by anti-$U$ quark of fourth generation
\cite{Q,I,lom,Khlopov:2006dk} (b) AC-leptons \cite{Khlopov:2006dk,5,FKS}, predicted in the
extension \cite{5} of standard model, based on the approach of
almost-commutative geometry \cite{bookAC}.  (c) Technileptons and anti-technibaryons
\cite{KK} in the framework of walking
technicolor models (WTC) \cite{Sannino:2004qp}. (d) Finally, stable
charged clusters $\bar u_5 \bar u_5 \bar u_5$ of (anti)quarks $\bar
u_5$ of 5th family can follow from the approach, unifying spins and
charges \cite{Norma}.

In the asymmetric case, corresponding to excess of -2 charge
species, $X^{--}$, as it was assumed for $(\bar U \bar U \bar
U)^{--}$ in the model of stable $U$-quark of a 4th generation, as
well as can take place for $(\bar u_5 \bar u_5 \bar u_5)^{--}$ in
the approach \cite{Norma} their positively charged partners
effectively annihilate in the early Universe. Such an asymmetric
case was realized in \cite{KK} in the framework of WTC, where it was
possible to relate the excess of negatively
charged anti-techni-baryons $(\bar U \bar U )^{--}$ and/or
technileptons $\zeta^{--}$ to the baryon asymmetry of the Universe.
The relationship between baryon asymmetry and excess of -2 charge stable species
is supported by sphaleron transitions at high temperatures and can be realized in all the models,
in which new stable species belong to non-trivial representations of electroweak SU(2) group.

 After it is formed
in the Standard Big Bang Nucleosynthesis (SBBN), $^4He$ screens the
$X^{--}$ charged particles in composite $(^4He^{++}X^{--})$ {\it
O-helium} ``atoms''
 \cite{I}.
 For different models of $X^{--}$ these "atoms" are also
called ANO-helium \cite{lom,Khlopov:2006dk}, Ole-helium
\cite{Khlopov:2006dk,FKS} or techni-O-helium \cite{KK}. We'll call
them all O-helium ($OHe$) in our further discussion, which follows
the guidelines of \cite{I2}.

In all these forms of O-helium, $X^{--}$ behaves either as lepton or
as specific "heavy quark cluster" with strongly suppressed hadronic
interaction. Therefore O-helium interaction with matter is
determined by nuclear interaction of $He$. These neutral primordial
nuclear interacting objects contribute to the modern dark matter
density and play the role of a nontrivial form of strongly
interacting dark matter \cite{Starkman,McGuire:2001qj}. The active
influence of this type of dark matter on nuclear transformations
seems to be incompatible with the expected dark matter properties.
However, it turns out that the considered scenario is not easily
ruled out \cite{I,FKS,KK,Khlopov:2008rp} and challenges the
experimental search for various forms of O-helium and its charged
constituents.

Here after a brief review of main features of OHe Universe we
concentrate on its effects in underground detectors. We present
following \cite{Bled09} a qualitative confirmation of the earlier guess \cite{I,I2,KK2} that
the positive results of dark matter searches in DAMA/NaI (see for
review \cite{Bernabei:2003za}) and DAMA/LIBRA \cite{Bernabei:2008yi}
experiments can be explained by effect of O-helium, resolving the controversy
between these data and negative results of other experimental
groups.

\section{O-helium Universe}

Following \cite{I,lom,Khlopov:2006dk,KK,I2} consider charge
asymmetric case, when excess of $X^{--}$ provides effective
suppression of positively charged species.

In the period $100\s \le t \le 300\s$  at $100 \keV\ge T \ge T_o=
I_{o}/27 \approx 60 \keV$, $^4He$ has already been formed in the
SBBN and virtually all free $X^{--}$ are trapped by $^4He$ in
O-helium ``atoms" $(^4He^{++} X^{--})$. Here the O-helium ionization
potential is\footnote{The account for charge distribution in $He$
nucleus leads to smaller value $I_o \approx 1.3 \MeV$
\cite{Pospelov}.} \beq I_{o} = Z_{x}^2 Z_{He}^2 \alpha^2 m_{He}/2
\approx 1.6 \MeV,\label{IO}\eeq where $\alpha$ is the fine structure
constant,$Z_{He}= 2$ and $Z_{x}= 2$ stands for the absolute value of
electric charge of $X^{--}$.  The size of these ``atoms" is
\cite{I,FKS} \beq R_{o} \sim 1/(Z_{x} Z_{He}\alpha m_{He}) \approx 2
\cdot 10^{-13} \cm \label{REHe} \eeq Here and further, if not
specified otherwise, we use the system of units $\hbar=c=k=1$.

O-helium, being an $\alpha$-particle with screened electric charge,
can catalyze nuclear transformations, which can influence primordial
light element abundance and cause primordial heavy element
formation. These effects need a special detailed and complicated
study. The arguments of \cite{I,FKS,KK} indicate that this model
does not lead to immediate contradictions with the observational
data.

Due to nuclear interactions of its helium constituent with nuclei in
the cosmic plasma, the O-helium gas is in thermal equilibrium with
plasma and radiation on the Radiation Dominance (RD) stage, while
the energy and momentum transfer from plasma is effective. The
radiation pressure acting on the plasma is then transferred to
density fluctuations of the O-helium gas and transforms them in
acoustic waves at scales up to the size of the horizon.

At temperature $T < T_{od} \approx 200 S^{2/3}_3\eV$ the energy and
momentum transfer from baryons to O-helium is not effective
\cite{I,KK} because $$n_B \sv (m_p/m_o) t < 1,$$ where $m_o$ is the
mass of the $OHe$ atom and $S_3= m_o/(1 \TeV)$. Here \beq \sigma
\approx \sigma_{o} \sim \pi R_{o}^2 \approx
10^{-25}\cm^2\label{sigOHe}, \eeq and $v = \sqrt{2T/m_p}$ is the
baryon thermal velocity. Then O-helium gas decouples from plasma. It
starts to dominate in the Universe after $t \sim 10^{12}\s$  at $T
\le T_{RM} \approx 1 \eV$ and O-helium ``atoms" play the main
dynamical role in the development of gravitational instability,
triggering the large scale structure formation. The composite nature
of O-helium determines the specifics of the corresponding dark
matter scenario.

At $T > T_{RM}$ the total mass of the $OHe$ gas with density $\rho_d
= (T_{RM}/T) \rho_{tot} $ is equal to
$$M=\frac{4 \pi}{3} \rho_d t^3 = \frac{4 \pi}{3} \frac{T_{RM}}{T} m_{Pl}
(\frac{m_{Pl}}{T})^2$$ within the cosmological horizon $l_h=t$. In
the period of decoupling $T = T_{od}$, this mass  depends strongly
on the O-helium mass $S_3$ and is given by \cite{KK}\beq M_{od} =
\frac{T_{RM}}{T_{od}} m_{Pl} (\frac{m_{Pl}}{T_{od}})^2 \approx 2
\cdot 10^{44} S^{-2}_3 \g = 10^{11} S^{-2}_3 M_{\odot}, \label{MEPm}
\eeq where $M_{\odot}$ is the solar mass. O-helium is formed only at
$T_{o}$ and its total mass within the cosmological horizon in the
period of its creation is $M_{o}=M_{od}(T_{od}/T_{o})^3 = 10^{37}
\g$.

On the RD stage before decoupling, the Jeans length $\lambda_J$ of
the $OHe$ gas was restricted from below by the propagation of sound
waves in plasma with a relativistic equation of state
$p=\epsilon/3$, being of the order of the cosmological horizon and
equal to $\lambda_J = l_h/\sqrt{3} = t/\sqrt{3}.$ After decoupling
at $T = T_{od}$, it falls down to $\lambda_J \sim v_o t,$ where $v_o
= \sqrt{2T_{od}/m_o}.$ Though after decoupling the Jeans mass in the
$OHe$ gas correspondingly falls down
$$M_J \sim v_o^3 M_{od}\sim 3 \cdot 10^{-14}M_{od},$$ one should
expect a strong suppression of fluctuations on scales $M<M_o$, as
well as adiabatic damping of sound waves in the RD plasma for scales
$M_o<M<M_{od}$. It can provide some suppression of small scale
structure in the considered model for all reasonable masses of
O-helium. The significance of this suppression and its effect on the
structure formation needs a special study in detailed numerical
simulations. In any case, it can not be as strong as the free
streaming suppression in ordinary Warm Dark Matter (WDM) scenarios,
but one can expect that qualitatively we deal with Warmer Than Cold
Dark Matter model.

Being decoupled from baryonic matter, the $OHe$ gas does not follow
the formation of baryonic astrophysical objects (stars, planets,
molecular clouds...) and forms dark matter halos of galaxies. It can
be easily seen that O-helium gas is collisionless for its number
density, saturating galactic dark matter. Taking the average density
of baryonic matter one can also find that the Galaxy as a whole is
transparent for O-helium in spite of its nuclear interaction. Only
individual baryonic objects like stars and planets are opaque for
it.

\section{Signatures of O-helium dark matter}
The composite nature of O-helium dark matter results in a number of
observable effects.
\subsection{Anomalous component of cosmic rays}
O-helium atoms can be destroyed in astrophysical processes, giving
rise to acceleration of free $X^{--}$ in the Galaxy.

O-helium can be ionized due to nuclear interaction with cosmic rays
\cite{I,I2}. Estimations \cite{I,Mayorov} show that for the number
density of cosmic rays $ n_{CR}=10^{-9}\cm^{-3}$ during the age of
Galaxy a fraction of about $10^{-6}$ of total amount of OHe is
disrupted irreversibly, since the inverse effect of recombination of
free $X^{--}$ is negligible. Near the Solar system it leads to
concentration of free $X^{--}$ $ n_{X}= 3 \cdot 10^{-10}S_3^{-1}
\cm^{-3}.$ After OHe destruction free $X^{--}$ have momentum of
order $p_{X} \cong \sqrt{2 \cdot M_{X} \cdot I_{o}} \cong 2 \GeV
S_3^{1/2}$ and velocity $v/c \cong 2 \cdot 10^{-3} S_3^{-1/2}$ and
due to effect of Solar modulation these particles initially can
hardly reach Earth \cite{KK2,Mayorov}. Their acceleration by Fermi
mechanism or by the collective acceleration forms power spectrum of
$X^{--}$ component at the level of $X/p \sim n_{X}/n_g = 3 \cdot
10^{-10}S_3^{-1},$ where $n_g \sim 1 \cm^{-3}$ is the density of
baryonic matter gas.

At the stage of red supergiant stars have the size $\sim 10^{15}
\cm$ and during the period of this stage$\sim 3 \cdot 10^{15} \s$,
up to $\sim 10^{-9}S_3^{-1}$ of O-helium atoms per nucleon can be
captured \cite{KK2,Mayorov}. In the Supernova explosion these OHe
atoms are disrupted in collisions with particles in the front of
shock wave and acceleration of free $X^{--}$ by regular mechanism
gives the corresponding fraction in cosmic rays.

If these mechanisms of $X^{--}$ acceleration are effective, the
anomalous low $Z/A$ component of $-2$ charged $X^{--}$ can be
present in cosmic rays at the level $X/p \sim n_{X}/n_g \sim
10^{-9}S_3^{-1},$ and be within the reach for PAMELA and AMS02
cosmic ray experiments.

In the framework of Walking Tachnicolor model the excess of both stable $X^{--}$
and $Y^{++}$ is possible \cite{KK2}, the latter being two-three orders of magnitude smaller, than the former.
It leads to the two-component composite dark matter scenario with the dominant OHe accompanied by a subdominant WIMP-like component of $(X^{--}Y^{++})$ bound systems. Technibaryons and technileptons can be metastable and decays of $X^{--}$ and $Y^{++}$ can provide explanation for anomalies, observed in high energy cosmic positron spectrum by PAMELA and in high energy electron spectrum by FERMI and ATIC.

\subsection{Positron annihilation and gamma lines in galactic
bulge} Inelastic interaction of O-helium with the matter in the
interstellar space and its de-excitation can give rise to radiation
in the range from few keV to few  MeV. In the galactic bulge with
radius $r_b \sim 1 \kpc$ the number density of O-helium can reach
the value $n_o\approx 3 \cdot 10^{-3}/S_3 \cm^{-3}$ and the
collision rate of O-helium in this central region was estimated in
\cite{I2}: $dN/dt=n_o^2 \sigma v_h 4 \pi r_b^3 /3 \approx 3 \cdot
10^{42}S_3^{-2} \s^{-1}$. At the velocity of $v_h \sim 3 \cdot 10^7
\cm/\s$ energy transfer in such collisions is $\Delta E \sim 1 \MeV
S_3$. These collisions can lead to excitation of O-helium. If 2S
level is excited, pair production dominates over two-photon channel
in the de-excitation by $E0$ transition and positron production with
the rate $3 \cdot 10^{42}S_3^{-2} \s^{-1}$ is not accompanied by
strong gamma signal. According to \cite{Finkbeiner:2007kk} this rate
of positron production for $S_3 \sim 1$ is sufficient to explain the
excess in positron annihilation line from bulge, measured by
INTEGRAL (see \cite{integral} for review and references). If $OHe$
levels with nonzero orbital momentum are excited, gamma lines should
be observed from transitions ($ n>m$) $E_{nm}= 1.598 \MeV (1/m^2
-1/n^2)$ (or from the similar transitions corresponding to the case
$I_o = 1.287 \MeV $) at the level $3 \cdot 10^{-4}S_3^{-2}(\cm^2 \s
\MeV ster)^{-1}$.

\section{O-helium in the terrestrial matter}
The evident consequence of the O-helium dark matter is its
inevitable presence in the terrestrial matter, which appears opaque
to O-helium and stores all its in-falling flux.

After they fall down terrestrial surface the in-falling $OHe$
particles are effectively slowed down due to elastic collisions with
matter. Then they drift, sinking down towards the center of the
Earth with velocity \beq V = \frac{g}{n \sigma v} \approx 80 S_3
A^{1/2} \cm/\s. \label{dif}\eeq Here $A \sim 30$ is the average
atomic weight in terrestrial surface matter, $n=2.4 \cdot 10^{24}/A$
is the number of terrestrial atomic nuclei, $\sigma v$ is the rate
of nuclear collisions and $g=980~ \cm/\s^2$.

Near the Earth's surface, the O-helium abundance is determined by
the equilibrium between the in-falling and down-drifting fluxes.

The in-falling O-helium flux from dark matter halo is
$$
  F=\frac{n_{0}}{8\pi}\cdot |\overline{V_{h}}+\overline{V_{E}}|,
$$
where $V_{h}$-speed of Solar System (220 km/s), $V_{E}$-speed of
Earth (29.5 km/s) and $n_{0}=3 \cdot 10^{-4} S_3^{-1} \cm^{-3}$ is the
local density of O-helium dark matter. Here, for qualitative estimation, we don't take into account velocity dispersion and distribution of particles in the incoming flux that can lead to significant effect.

At a depth $L$ below the Earth's surface, the drift timescale is
$t_{dr} \sim L/V$, where $V \sim 400 S_3 \cm/\s$ is given by
Eq.~(\ref{dif}). It means that the change of the incoming flux,
caused by the motion of the Earth along its orbit, should lead at
the depth $L \sim 10^5 \cm$ to the corresponding change in the
equilibrium underground concentration of $OHe$ on the timescale
$t_{dr} \approx 2.5 \cdot 10^2 S_3^{-1}\s$.

The equilibrium concentration, which is established in the matter of
underground detectors at this timescale, is given by
\begin{equation}
    n_{oE}=\frac{2\pi \cdot F}{V} = n_{0}\frac{n \sigma v}{4g} \cdot
    |\overline{V_{h}}+\overline{V_{E}}|,
\end{equation}
where, with account for $V_{h} > V_{E}$, relative velocity can be
expressed as
$$
    |\overline{V_{o}}|=\sqrt{(\overline{V_{h}}+\overline{V_{E}})^{2}}=\sqrt{V_{h}^2+V_{E}^2+V_{h}V_{E}sin(\theta)} \simeq
$$
$$
\simeq V_{h}\sqrt{1+\frac{V_{E}}{V_{h}}sin(\theta)}\sim
V_{h}(1+\frac{1}{2}\frac{V_{E}}{V_{h}}sin(\theta)).
$$
Here $\theta=\omega (t-t_0)$ with $\omega = 2\pi/T$, $T=1yr$ and
$t_0$ is the phase. Then the concentration takes the form
\begin{equation}
    n_{oE}=n_{oE}^{(1)}+n_{oE}^{(2)}\cdot sin(\omega (t-t_0))
    \label{noE}
\end{equation}

So, there are two parts of the signal: constant and annual
modulation, as it is expected in the strategy of dark matter search
in DAMA experiment \cite{Bernabei:2008yi}.

Such neutral $(^4He^{++}X^{--})$ ``atoms" may provide a catalysis of
cold nuclear reactions in ordinary matter (much more effectively
than muon catalysis). This effect needs a special and thorough
investigation. On the other hand, $X^{--}$ capture by nuclei,
heavier than helium, can lead to production of anomalous isotopes,
but the arguments, presented in \cite{I,FKS,KK} indicate that their
abundance should be below the experimental upper limits.

It should be noted that the nuclear cross section of the O-helium
interaction with matter escapes the severe constraints
\cite{McGuire:2001qj} on strongly interacting dark matter particles
(SIMPs) \cite{Starkman,McGuire:2001qj} imposed by the XQC experiment
\cite{XQC}. Therefore, a special strategy of direct O-helium  search
is needed, as it was proposed in \cite{Belotsky:2006fa}.

In underground detectors, $OHe$ ``atoms'' are slowed down to thermal
energies and give rise to energy transfer $\sim 2.5 \cdot 10^{-4}
\eV A/S_3$, far below the threshold for direct dark matter
detection. It makes this form of dark matter insensitive to the
severe CDMS constraints \cite{Akerib:2005kh}. However, in $OHe$ reactions
with the matter of underground detectors  can lead to observable effects.
Following earlier guess \cite{I,I2,KK2,Bled08} it was shown in \cite{Bled09}
that such reactions in NaI can explain the results of DAMA/NaI and DAMA/LIBRA experiments.
\section{Low energy bound state of O-helium with nuclei}

The explanation \cite{Bled09} is based on the idea that OHe,
slowed down in the matter of DAMA/NaI or DAMA/LIBRA detector, can form a few keV bound state with
nucleus, in which OHe is situated \textbf{beyond} the nucleus.
Therefore the positive result of this experiment is explained by reaction
\begin{equation}
A+(^4He^{++}X^{--}) \rightarrow [A(^4He^{++}X^{--})]+\gamma
\label{HeEAZ}
\end{equation}
with sodium and/or iodine.
In detectors with different chemical content such level may not exist at all, or has other value of energy, making the the comparison with DAMA results a nontrivial task.

\subsection{Low energy bound state of O-helium with nuclei}

The approach of \cite{Bled09} assumes the following picture: at the distances larger, than its size,
OHe is neutral and it feels only Yukawa exponential tail of nuclear attraction,
due to scalar-isoscalar nuclear potential. It should be noted that scalar-isoscalar
nature of He nucleus excludes its nuclear interaction due to $\pi$ or $\rho$ meson exchange,
so that the main role in its nuclear interaction outside the nucleus plays $\sigma$ meson exchange,
on which nuclear physics data are not very definite. When the distance from the surface of nucleus becomes
smaller than the size of OHe, the mutual attraction of nucleus and OHe is changed by dipole Coulomb repulsion. Inside the nucleus strong nuclear attraction takes place. In the result the spherically symmetric potential appears,given by
\begin{equation}
U=-\frac{A_{He} A g^2 exp(-\mu r)}{r} + \frac{Z_{He} Z e^2 r_o \cdot F(r)}{r^2}.
\label{epot}
\end{equation}
Here $A_{He}=4$, $Z_{He}=2$ are atomic weight and charge of helium, $A$ and $Z$ are respectively atomic weight and charge of nucleus, $\mu$ and $g^2$ are the mass and coupling of scalar-isoscalar meson - mediator of nuclear attraction, $r_o$ is the size of OHe and $F(r)$ is its electromagnetic formfactor, which strongly suppresses the strength of dipole electromagnetic interaction outside the OHe "atom".

Schrodinger equation for this system is reduced
(taking apart the equation for the center of mass) to the equation of relative motion for the reduced mass.

In the case of orbital momentum \emph{l}=0 the wave functions depend only on \emph{r}.

\begin{figure}
         \includegraphics[width=4in]{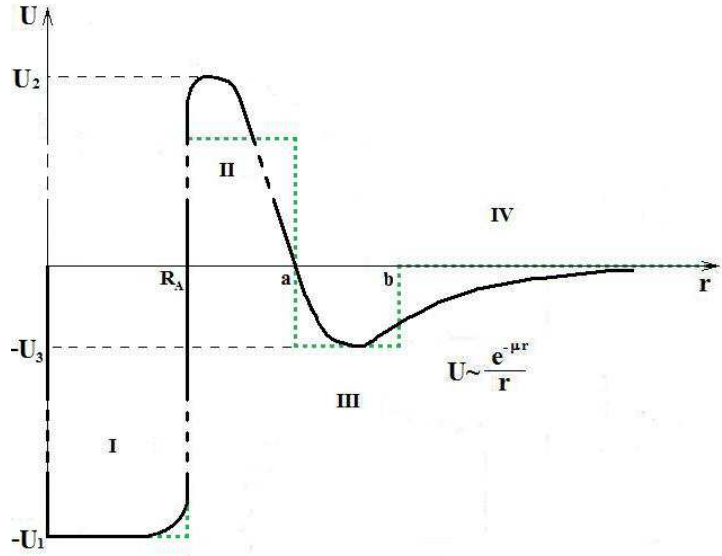}\\
        \caption{The approximation of rectangular well for potential of OHe-nucleus system.}\label{pic23}
\end{figure}

To simplify the solution of Schrodinger equation the potential (\ref{epot}) was approximated in \cite{Bled09}
by a rectangular potential that
consists of a deep potential well within the
radius of nucleus $R_A$, of a rectangular dipole Coulomb potential barrier outside its surface up to the
 radial layer $a=R_A+r_o$, where it is suppressed by the OHe atom formfactor, and of the outer potential well of the width $\sim 1/\mu$, formed by the tail of Yukawa nuclear interaction. It leads to the approximate potential \cite{Bled09}, presented on Fig. \ref{pic23}.

Solutions of Schrodinger
equation for each of the four regions, indicated on Fig. \ref{pic23}, are given in textbooks (see e.g.\cite{LL3}) and
 their sewing determines the condition, under which a low-energy  OHe-nucleus bound state appears in the region III.

\begin{figure}
           \includegraphics[width=4in]{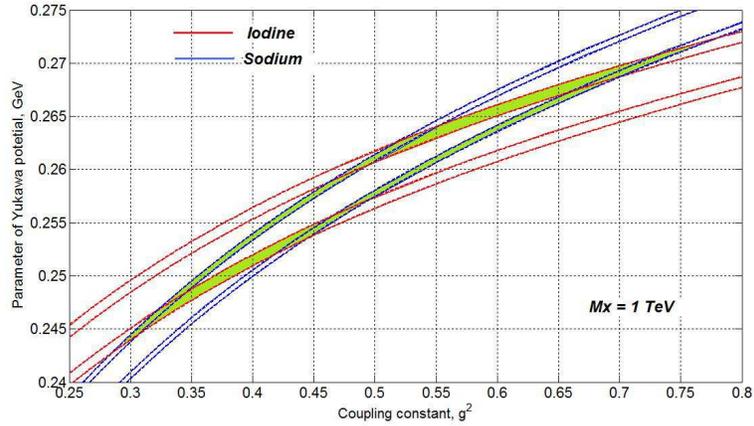}\\
        \caption{The region of parameters $\mu$ and $g^2$, for which Na and I have a level in the interval 2-6 keV. For each nucleus two narrow strips determine the region of parameters, at which the bound system of this element with OHe has a level in 2-6 keV energy range. The outer line of strip corresponds to the level of 6 keV and the internal line to the level of 2 keV. The region of intersection of strips correspond to existence of 2-6 keV levels in both OHe-Na and OHe-I systems, while the piece of strip between strips of other nucleus corresponds to the case, when OHe bound state with this nucleus has 2-6 keV level, while the binding energy of OHe with the other nuclei is less than 2 keV by absolute value.}\label{NaI}
   \end{figure}

The energy of this bound state and its existence strongly depend on the parameters $\mu$ and $g^2$ of nuclear potential (\ref{epot}). On the Fig. \ref{NaI} the region of these parameters, giving 2-6 keV energy level in OHe bound states with sodium and iodine are presented. In these calculations \cite{Bled09} the mass of OHe was taken equal to $m_o=1 TeV$.

The rate of radiative capture of OHe by nuclei should be accurately calculated with the use of exact form of wave functions, obtained for the OHe-nucleus bound state. This work is now in progress. One can use the analogy with the radiative capture of neutron by proton with the following corrections:
 \begin{itemize}
\item
  There is only E1 transition in the case of OHe capture.
\item
  The reduced masses of n-p and OHe-nucleus systems are different
\item
  The existence of dipole Coulomb barrier leads to a suppression of the cross section of OHe radiative capture.
\end{itemize}
 With the account for these effects our first estimations give the rate of OHe radiative capture, reproducing the level of signal, detected by DAMA.

Formation of OHe-nucleus bound system leads to energy release of its binding energy, detected as ionization signal in DAMA experiment. In the context of the approach \cite{Bled09} the existence of annual modulations of this signal in the range 2-6 keV and absence of such effect at energies above 6 keV means that binding energy of Na-OHe and I-OHe systems should not exceed 6 keV, being in the range 2-6 keV for at least one of these elements. These conditions were taken into account for determination of nuclear parameters, at which the result of DAMA can be reproduced. At these values of $\mu$ and $g^2$ energy of OHe binding with other nuclei can strongly differ from 2-6 keV. In particular, energy release at the formation of OHe bound state with thallium can be larger than 6 keV. However, assuming $A^{-2}$ dependence for the cross section of radiative capture of nuclei by OHe and taking into account that thallium content in DAMA detector is 3 orders of magnitude smaller, than NaI, such signal is to be below the experimental errors.

It should be noted that the results of DAMA experiment exhibit also absence of annual modulations at the energy of MeV-tens MeV. Energy release in this range should take place, if OHe-nucleus system comes to the deep level inside the nucleus (in the region I of Fig. \ref{pic23}). This transition implies tunneling through dipole Coulomb barrier and is suppressed below the experimental limits.

\subsection{Energy levels in other nuclei}
\begin{figure}
            \includegraphics[width=4in]{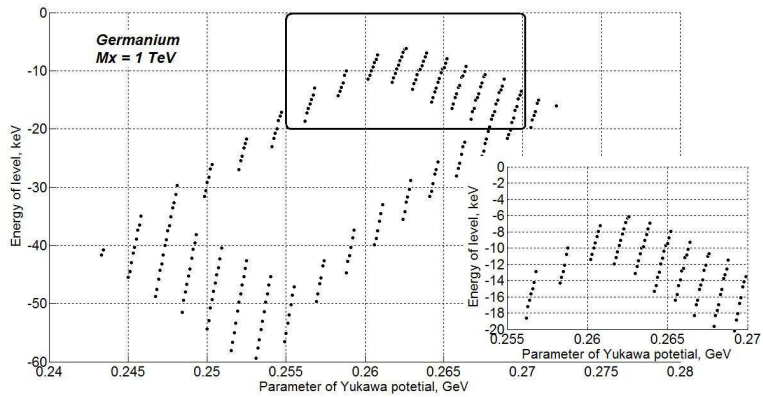}\\
        \caption{Energy levels in OHe bound system with germanium. The range of energies close to energy release in DAMA experiment is blown up to demonstrate that even in this range there is no formal intersection with DAMA results. }\label{Ge}
    \end{figure}

For the chosen range of nuclear parameters, reproducing the results of DAMA/NaI and DAMA/LIBRA, the binding energy of OHe-nucleus states in nuclei, corresponding to chemical composition of set-ups in other experiments were calculated in \cite{Bled09}. The results of such calculation for germanium, corresponding to the CDMS detector of experiment, are presented on Fig. \ref{Ge}.
For all the parameters, reproducing results of DAMA experiment the predicted energy level of OHe-germanium bound state is beyond the range 2-6 keV, being dominantly in the range of tens - few-tens keV by absolute value. It makes elusive a possibility to test DAMA results by search for ionization signal in the same range 2-6 keV in other set-ups with content that differs from Na and I. In particular, our approach naturally predicts absence of ionization signal in the range 2-6 keV in accordance with the recent results of CDMS \cite{Kamaev:2009gp}.

\begin{figure}
            \includegraphics[width=4in]{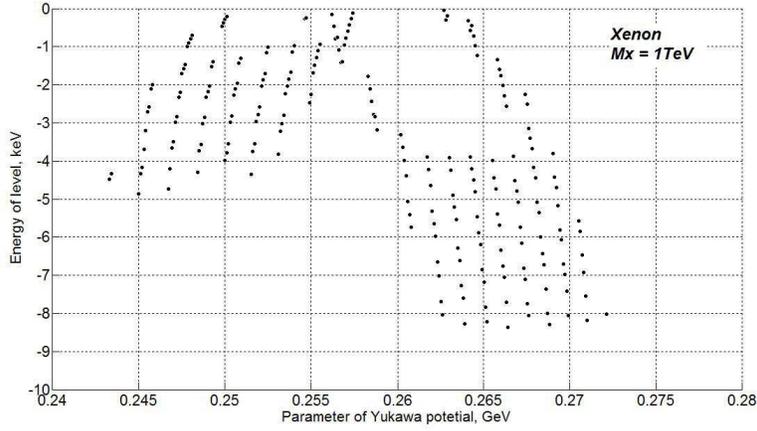}\\
        \caption{Energy levels in OHe bound system with xenon.}\label{Xe}
    \end{figure}

\begin{figure}
            \includegraphics[width=4in]{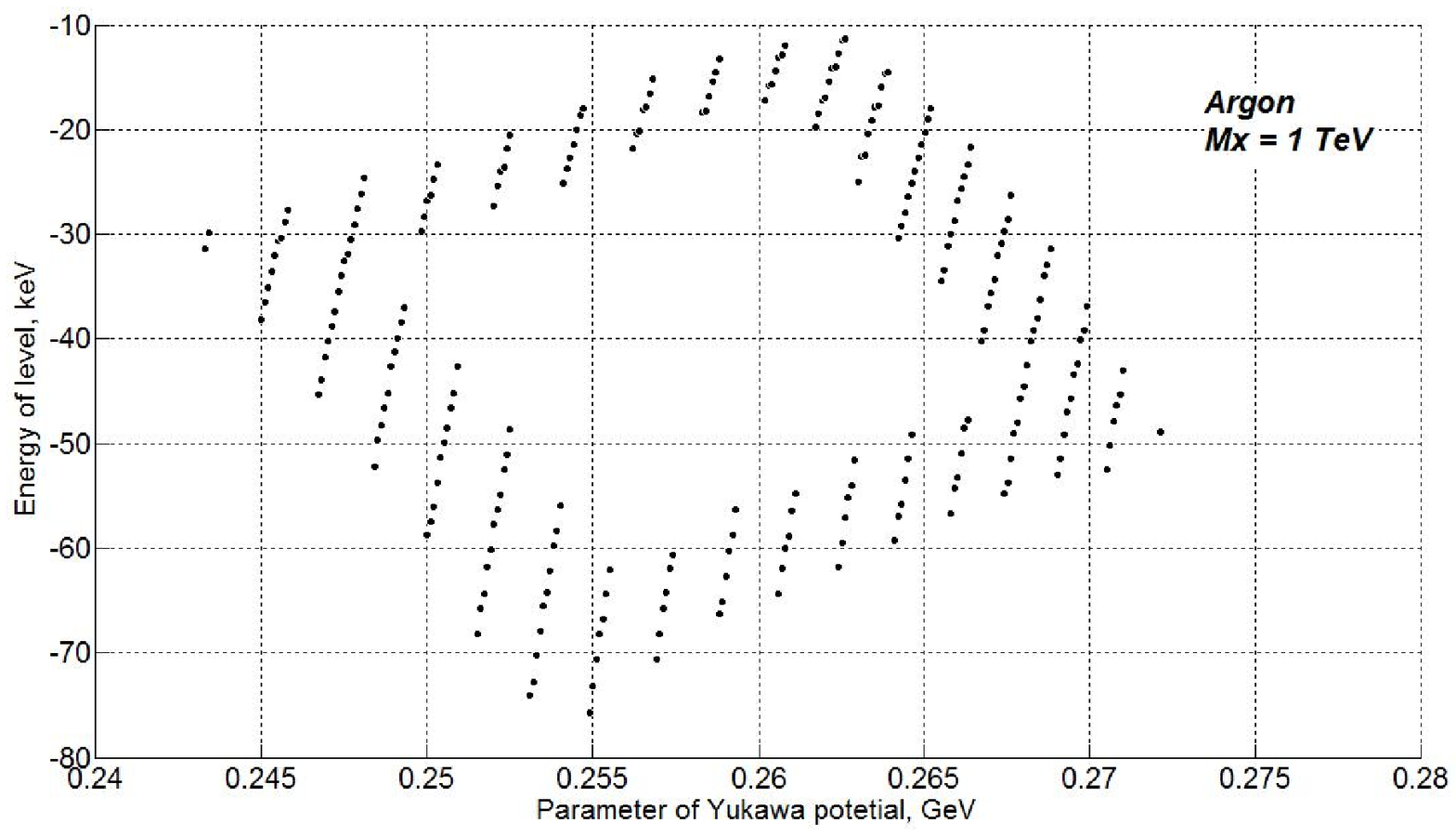}\\
        \caption{Energy levels in OHe bound system with argon.}\label{Ar}
    \end{figure}

\begin{figure}
            \includegraphics[width=4in]{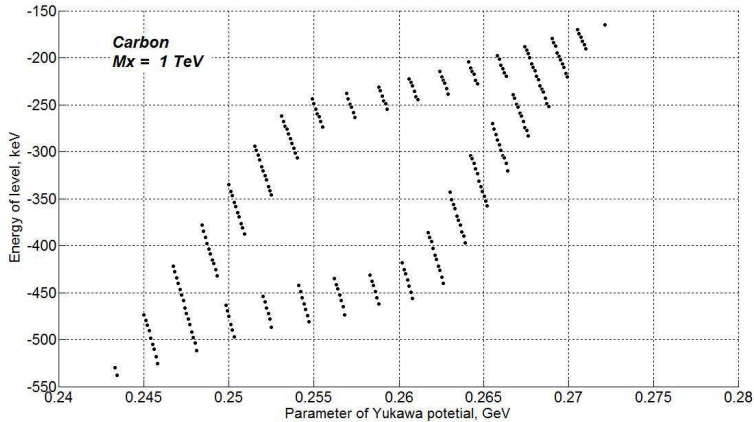}\\
        \caption{Energy levels in OHe bound system with carbon.}\label{C}
    \end{figure}

There were also calculated the energies of bound states of OHe with xenon (Fig. \ref{Xe}), argon (Fig. \ref{Ar}) and carbon (Fig. \ref{C}).

With the growth of the mass of O-helium the reduced mass of OHe-nucleus system slightly grows, approaching with higher accuracy the mass of nucleus. It extends a bit the range of nuclear parameters $\mu$ and $g^2$, at which the binding energy of OHe with sodium and/or iodine is within the range 2-6 keV. Qualitatively, the predictions for superheavy O-helium are similar to the case of $S_3=1$.
\section{Conclusions}

To conclude, the existence of heavy stable charged particles may not only be compatible with the experimental constraints but even lead to composite dark matter scenario of nuclear interacting Warmer than Cold Dark Matter. This new form of dark matter can provide explanation of excess of positron annihilation line radiation, observed by INTEGRAL in the galactic bulge. The search for stable -2 charge component of cosmic rays is challenging for PAMELA and AMS02 experiments. Decays of heavy charged constituents of composite dark matter can provide explanation for anomalies in spectra of cosmic high energy positrons and electrons, observed by PAMELA, FERMI and ATIC. In the context of our approach search for heavy stable charged quarks and leptons at LHC acquires the significance of experimental probe for components of cosmological composite dark matter.

The results of dark matter search in experiments
DAMA/NaI and DAMA/LIBRA can be explained in the framework of
our scenario without contradiction with negative
results of other groups. This scenario can be realized in different
frameworks, in particular, in the extensions of Standard Model, based on the approach of almost commutative geometry, in the model of stable quarks of 4th generation that can be naturally embedded in the heterotic superstring phenomenology, in the models of stable technileptons and/or techniquarks, following from Minimal Walking Technicolor model or in
the approach unifying spin and charges. Our approach contains distinct
features, by which the present explanation can be distinguished from
other recent approaches to this problem \cite{Edward} (see also
review and more references in \cite{Gelmini} as well as in the corresponding contributions to the present Proceedings).

The proposed explanation is based on the mechanism of low energy binding of OHe with nuclei.
Within the uncertainty of nuclear physics parameters there exists a range at which OHe
binding energy with sodium and/or iodine is in the interval 2-6 keV. Radiative capture of OHe to this bound state leads to the corresponding energy release observed as an ionization signal
in DAMA detector.

OHe concentration in the matter of underground detectors is determined by the equilibrium between the incoming cosmic flux of OHe and diffusion towards the center of Earth. It is rapidly adjusted and follows the
change in this flux with the relaxation time of few
minutes. Therefore the rate of radiative capture of OHe should experience annual modulations reflected in annula modulations of the ionization signal from these reactions.


An inevitable consequence of the proposed explanation is appearance
in the matter of DAMA/NaI or DAMA/LIBRA detector anomalous
superheavy isotopes of sodium and/or iodine,
having the mass roughly by $m_o$ larger, than ordinary isotopes of
these elements. If the atoms of these anomalous isotopes are not
completely ionized, their mobility is determined by atomic cross
sections and becomes about 9 orders of magnitude smaller, than for
O-helium. It provides their conservation in the matter of detector. Therefore mass-spectroscopic
analysis of this matter can provide additional test for the O-helium
nature of DAMA signal. Methods of such analysis should take into account
the fragile nature of OHe-Na bound states, since their binding energy is only few keV.

With the account for high sensitivity of the numerical results to the values of nuclear parameters
and for the approximations, made in the calculations, the presented results can be considered
only as an illustration of the possibility to explain puzzles of dark matter search in
the framework of composite dark matter scenario. An interesting feature of this explanation is a conclusion that the ionization signal expected in detectors
with the content, different from NaI, can be dominantly in the energy range beyond 2-6 keV.
Therefore test of results of DAMA/NaI and DAMA/LIBRA experiments by other experimental groups can become a very nontrivial task.

Our results show that the ionization signal, detected by DAMA, may be absent in detectors containing light elements. In particular, there is predicted no low-energy binding of OHe with $^3He$ and correspondingly no ionization signal in keV range in the designed $^3He$ dark matter detectors. Therefore development of experimental methods of dark matter detection will extend the possibilities to test hypothesis of composite dark matter.

The presented approach sheds new light on the physical nature of dark matter. Specific properties of composite dark matter and its constituents are challenging for their experimental search. OHe interaction with matter is an important aspect of these studies. In this context positive result of DAMA/NaI and DAMA/LIBRA experiments may be a signature for exciting phenomena of O-helium nuclear physics.


\begin{theacknowledgments}
  I express my gratitude to K.M. Belotsky, C. Kouvaris, A.G. Mayorov and E. Yu. Soldatov for collaboration in obtaining the presented results and to Jean Pierre Gazeau for discussions.
\end{theacknowledgments}



\bibliographystyle{aipproc}   

\bibliography{sample}

\IfFileExists{\jobname.bbl}{}
 {\typeout{}
  \typeout{******************************************}
  \typeout{** Please run "bibtex \jobname" to optain}
  \typeout{** the bibliography and then re-run LaTeX}
  \typeout{** twice to fix the references!}
  \typeout{******************************************}
  \typeout{}
 }

\end{document}